\documentclass{article}

\usepackage[preprint]{neurips_2026}

\usepackage[utf8]{inputenc}
\usepackage[T1]{fontenc}
\usepackage{amsmath,amssymb}
\usepackage{graphicx}
\usepackage{booktabs}
\usepackage[hidelinks]{hyperref}
\usepackage{xcolor}
\usepackage{listings}
\usepackage{natbib}
\usepackage{algorithm}
\usepackage{algpseudocode}
\usepackage{float}
\usepackage{multirow}
\usepackage{tikz}
\usetikzlibrary{positioning, shapes.geometric, arrows.meta, calc}

\newcommand{\metric}[1]{\textsc{#1}}

\title{Compiled AI: Deterministic Code Generation for\\LLM-Based Workflow Automation}

\author{%
  Geert Trooskens\textsuperscript{*,1} \quad
  Anmol Sharma\textsuperscript{*,1} \quad
  Aaron Karlsberg\textsuperscript{1} \quad
  Lamara De Brouwer\textsuperscript{1} \\[0.3em]
  Max Van Puyvelde\textsuperscript{2} \quad
  Matthew Young\textsuperscript{1} \quad
  John Thickstun\textsuperscript{3} \\[0.3em]
  Gil Alterovitz\textsuperscript{4} \quad
  Walter A. De Brouwer\textsuperscript{2} \\[1em]
  \textsuperscript{1}XY.AI Labs, Palo Alto, CA \quad
  \textsuperscript{2}Stanford University School of Medicine, Stanford, CA \\
  \textsuperscript{3}Cornell University, Department of Computer Science, Ithaca, NY \\
  \textsuperscript{4}Brigham and Women's Hospital / Harvard Medical School, Boston, MA%
}

\begin{document}
\raggedbottom

\maketitle
\renewcommand{\thefootnote}{*}%
\phantomsection
\footnotetext{Equal contribution. Correspondence to: \texttt{geert@xy.ai}.}%
\renewcommand{\thefootnote}{\arabic{footnote}}

\begin{abstract}

We study \emph{compiled AI}, a paradigm in which large language models generate executable code artifacts during a compilation phase, after which workflow orchestration executes deterministically with the LLM removed from the control flow. Branching, tool selection, error handling, and retry logic all run as static code with zero LLM calls. Where a step requires semantic understanding such as OCR cleanup or clinical note extraction, the compiled artifact invokes the LLM as a bounded tool call whose schema and trigger are fixed at compile time, so runtime inference is confined to narrow subtasks rather than open-ended orchestration. This paradigm has antecedents in prior work on declarative pipeline optimization (DSPy) and hybrid neural-symbolic planning (LLM+P); our contribution is a systems-oriented study of its application to high-stakes enterprise workflows, with particular emphasis on healthcare settings where reliability and auditability are critical.

By constraining generation to narrow business-logic functions embedded in validated templates, compiled AI trades runtime flexibility for predictability, auditability, cost efficiency, and reduced security exposure. We introduce (i) a system architecture for constrained LLM-based code generation, (ii) a four-stage generation-and-validation pipeline that converts probabilistic model output into production-ready code artifacts, and (iii) an evaluation framework measuring operational metrics including token amortization, determinism, reliability, security, and cost.

We evaluate on two task types: function-calling (BFCL, $n$=400) and document intelligence (DocILE, $n$=5,680 invoices). On function-calling, where every step is structured and no runtime inference is required, compiled AI achieves 96\% task completion with zero runtime tokens, breaking even with runtime inference at approximately 17 transactions and reducing token consumption by 57$\times$ at 1,000 transactions. On document intelligence, where extraction steps require semantic understanding, our Code Factory variant compiles orchestration deterministically and invokes the LLM as a bounded tool for field extraction, matching Direct LLM on key field extraction (KILE: 80.0\%) while achieving the highest line item recognition accuracy (LIR: 80.4\%). Security evaluation across 135 test cases demonstrates 96.7\% accuracy on prompt injection detection and 87.5\% on static code safety analysis with zero false positives.

\end{abstract}

\section{Introduction}
\label{sec:introduction}

Large language models (LLMs) are increasingly deployed to automate enterprise workflows, evolving from question-answering systems toward autonomous agent architectures \citep{autogen2024, crewai2024, langchain2022}. While these approaches demonstrate flexibility, they rely on repeated model invocation during execution, leading to high token consumption, variable latency, and non-deterministic behavior.

The idea of using LLMs as compilers rather than interpreters is not new. DSPy \citep{khattab2023dspy} compiles declarative LLM calls into optimized pipelines, achieving 25--65\% improvement over prompt engineering. LLM+P \citep{liu2023llmp} translates natural language to PDDL, then uses classical planners to produce optimal solutions. The LLM handles intent understanding, the deterministic solver handles execution. Text-to-SQL systems instantiate the same pattern: an LLM generates a query once, and the database executes it deterministically at scale. Our contribution is to study this paradigm as a first-class production systems concern, evaluating it with operational metrics suited to enterprise deployment, and demonstrating its particular value in healthcare settings where reliability and auditability are regulatory requirements.

Empirical studies reveal persistent reliability challenges in runtime agent systems. \citet{cemri2025multiagent} found that 79\% of multi-agent failures stem from specification and coordination issues rather than infrastructure. Salesforce's CRMArena-Pro benchmark shows agent success rates degrading from 58\% in single-turn to 35\% in multi-turn interactions \citep{salesforce2025agents}. Non-determinism persists even at temperature=0: \citet{atil2024nondeterminism} found accuracy varies up to 15\% across runs, and \citet{ouyang2023nondeterminism} demonstrated 18--75\% output variance due to architectural factors including Mixture-of-Experts routing. These limitations are especially problematic in healthcare, where prior authorization workflows, clinical document processing, and billing automation require determinism, auditability, and compliance with HIPAA and CMS decision timeframes.

This paper investigates compiled AI as a design point optimized for well-specified, high-volume, compliance-sensitive workflows. The key observation is that many enterprise workflows require intelligence to \emph{design} but not to \emph{execute} repeatedly. Once business logic is specified and validated, repeated execution can be handled by conventional software infrastructure, with full observability enabling targeted refinement of compiled artifacts over time. Rather than interpreting natural language for orchestration at runtime, an LLM is constrained to generate narrow business-logic functions within pre-validated templates. These artifacts pass a multi-stage validation pipeline (security analysis, syntactic verification, execution testing, output accuracy) before deployment as static code with predictable orchestration latency, deterministic control flow, and amortized inference cost.

\paragraph{Definition (Compiled AI).} A workflow system satisfying three properties: (1) \emph{one-time LLM invocation}, where the model runs once at generation time, not at transaction time; (2) \emph{zero LLM invocations in the control plane}, where orchestration runs as static code and any runtime LLM calls are confined to bounded tool invocations with schemas fixed at compile time; and (3) \emph{mandatory multi-stage validation}, where every artifact passes security, syntax, execution, and accuracy checks before deployment. The comparison below situates compiled AI relative to runtime agent systems and DSPy.

\begin{center}
\small
\setlength{\tabcolsep}{6pt}
\begin{tabular}{lccc}
\toprule
\textbf{System} & \textbf{Runtime LLM?} & \textbf{Deterministic?} & \textbf{Validation required?} \\
\midrule
Agents (AutoGen, LangChain) & Yes & No  & No \\
DSPy                        & Partial & Partial & No \\
\textbf{Compiled AI}        & \textbf{Bounded only} & \textbf{Yes} & \textbf{Yes} \\
\bottomrule
\end{tabular}
\end{center}

\paragraph{Contributions.} This paper makes three contributions: (1) a system architecture for constrained LLM-based code generation supporting bounded runtime tool calls and producing bounded business-logic functions in durable workflow templates (Section~\ref{sec:architecture}); (2) a four-stage generation-and-validation pipeline converting probabilistic model output into production-ready code artifacts (Section~\ref{sec:architecture}); and (3) an evaluation framework measuring operational viability (token amortization, determinism, reliability, validation effectiveness, and cost) evaluated across two distinct task types (Sections~\ref{sec:framework}--\ref{sec:results}).

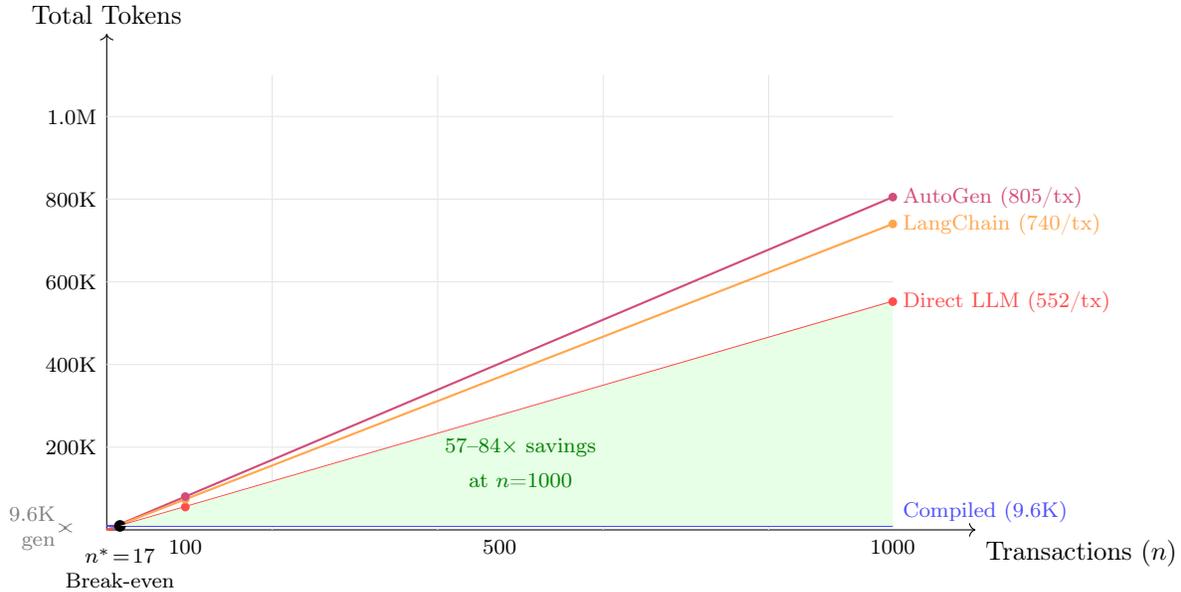
\begin{figure}[t]
\centering
\begin{tikzpicture}[scale=1.1]
    \draw[->] (0,0) -- (0,6) node[above] {\small Total Tokens};
    \draw[->] (0,0) -- (10.5,0) node[below right] {\small Transactions ($n$)};
    \foreach \x in {2,4,6,8} { \draw[gray!20] (\x,0) -- (\x,5.5); }
    \foreach \y in {1,2,3,4,5} { \draw[gray!20] (0,\y) -- (9.5,\y); }
    \node[left, font=\scriptsize] at (0,1) {200K};
    \node[left, font=\scriptsize] at (0,2) {400K};
    \node[left, font=\scriptsize] at (0,3) {600K};
    \node[left, font=\scriptsize] at (0,4) {800K};
    \node[left, font=\scriptsize] at (0,5) {1.0M};
    \draw[thick, purple!70] (0,0) -- (9.5,4.025);
    \node[purple!70, right, font=\scriptsize] at (9.5, 4.025) {AutoGen (805/tx)};
    \draw[thick, orange!70] (0,0) -- (9.5,3.70);
    \node[orange!70, right, font=\scriptsize] at (9.5, 3.70) {LangChain (740/tx)};
    \draw[thick, red!70] (0,0) -- (9.5,2.76);
    \node[red!70, right, font=\scriptsize] at (9.5, 2.76) {Direct LLM (552/tx)};
    \draw[thick, blue!70] (0,0.048) -- (9.5,0.048);
    \node[blue!70, right, font=\scriptsize] at (9.5, 0.22) {Compiled (9.6K)};
    \draw[<->, gray] (-0.5,0) -- (-0.5,0.048);
    \node[gray, left, font=\scriptsize, align=right] at (-0.5,0.024) {9.6K\\gen};
    \draw[dashed, gray] (0.16,0) -- (0.16,0.048);
    \node[below, font=\scriptsize] at (0.16,-0.1) {$n^*\!=\!17$};
    \fill[black] (0.16,0.048) circle (2pt);
    \node[below, font=\scriptsize] at (0.16,-0.4) {Break-even};
    \fill[green!10] (0.16,0.048) -- (9.5,0.048) -- (9.5,2.76) -- (0.16,0.048);
    \node[green!50!black, font=\scriptsize] at (5,1.0) {57--84$\times$ savings};
    \node[green!50!black, font=\scriptsize] at (5,0.6) {at $n$=1000};
    \node[below, font=\scriptsize] at (0.95,0) {100};
    \node[below, font=\scriptsize] at (4.75,0) {500};
    \node[below, font=\scriptsize] at (9.5,0) {1000};
    \fill[red!70] (0.95,0.276) circle (1.5pt);
    \fill[red!70] (9.5,2.76) circle (1.5pt);
    \fill[orange!70] (0.95,0.37) circle (1.5pt);
    \fill[orange!70] (9.5,3.70) circle (1.5pt);
    \fill[purple!70] (0.95,0.4025) circle (1.5pt);
    \fill[purple!70] (9.5,4.025) circle (1.5pt);
\end{tikzpicture}
\caption{Token consumption comparison across baselines (BFCL, n=400). Compiled AI incurs a one-time 9,600-token generation cost, then executes at zero marginal cost. Break-even vs.\ Direct LLM at $n^*\approx17$; at 1,000 transactions, compiled AI reduces consumption by 57$\times$ vs.\ Direct LLM and 84$\times$ vs.\ AutoGen.}
\label{fig:economics}
\end{figure}

\section{Related Work}
\label{sec:related-work}

\subsection{The Compilation Paradigm}
\label{sec:related-compilation}

We position compiled AI within the broader shift toward compound AI systems~\citep{zaharia2024shift}, where capabilities are delivered through composed components rather than single models. Within this framing, three threads of prior work motivate the paradigm.

The most directly relevant validates compilation as a productive paradigm for LLM systems. DSPy \citep{khattab2023dspy} compiles declarative LLM calls into optimized pipelines, achieving 25--65\% improvement over prompt engineering and enabling small 770M-parameter models to compete with GPT-3.5 on several tasks. LLM+P \citep{liu2023llmp} translates natural language to PDDL and delegates execution to classical planners; LLMs alone fail to produce even feasible plans, but the hybrid approach succeeds. WorkflowLLM \citep{fan2024workflowllm} demonstrates one-shot workflow learning with zero-shot transfer to unseen APIs. FlowMind~\citep{zeng2023flowmind} generates workflows that prevent runtime hallucinations via structured intermediate representations.

A second thread concerns structured generation and tool invocation. Tool-augmented LLMs~\citep{schick2023toolformer} and program-aided approaches~\citep{gao2023pal, chen2023pot, li2024chainofcode} established the broader pattern of LLMs invoking external functions or generating code as a reasoning step; compiled AI fixes the tool selection and schema at compile time so that runtime invocation is bounded by a predetermined contract. AlphaCodium \citep{ridnik2024alphacodium} shows that structured multi-stage flows with verification improve GPT-4 accuracy from 19\% to 44\%. MetaGPT's Standard Operating Procedures \citep{hong2024metagpt} are essentially compiled workflows: deterministic procedures reducing cascading hallucinations by constraining agent behavior.

Concurrent work by \citet{blueprint2025} also separates deterministic workflow structure from LLM-driven steps, evaluated on $\tau$-bench. We share the deterministic-orchestration intuition but differ on three points: (i) Blueprint keeps the LLM in every per-step decision, whereas compiled AI removes the LLM from execution entirely except where a step explicitly requires semantic understanding (and even there, invocation is bounded by a compile-time schema); (ii) we evaluate operational metrics including token amortization, determinism, validation pipeline yield, and security, rather than task completion alone; (iii) we evaluate on structured (BFCL) and noisy semantic (DocILE) inputs rather than $\tau$-bench's agent-style scenarios. More broadly, our work treats \emph{production viability} as a first-class concern, focusing on operational properties (token economics, determinism, auditability, cost at scale) that determine enterprise deployability in high-stakes domains where reliability is a regulatory requirement.

\subsection{Agent Frameworks and Runtime Orchestration}
\label{sec:related-frameworks}

A separate line of work develops runtime agent frameworks. LangChain \citep{langchain2022} and CrewAI \citep{crewai2024} provide composable abstractions for chained LLM calls and multi-agent coordination; AutoGen \citep{autogen2024} formalizes conversational multi-agent patterns; tree-search-based agents such as LATS~\citep{zhou2024lats} extend this with structured exploration. These frameworks treat the LLM as a runtime interpreter, invoking the model on every transaction, and rely on durable-execution infrastructure such as Temporal \citep{temporal2024} for state management. We use LangChain and AutoGen as runtime baselines to quantify the cost of keeping the LLM in the per-transaction loop. Compiled AI inverts this pattern: the LLM compiles a workflow once, and the resulting artifact runs on the same durable-execution substrate without further model calls.

\subsection{The Reliability Crisis in LLM Agents}
\label{sec:related-reliability}

A substantial body of empirical work motivates moving the LLM out of the orchestration loop. \citet{cemri2025multiagent} analyzed 1,600+ annotated traces across seven major frameworks, identifying specification failures (41.8\%) and inter-agent coordination failures (36.9\%) as the dominant failure modes. 79\% of failures are not infrastructure problems. \citet{pan2025measuring} surveyed 306 practitioners across 26 domains and found 68\% limit agents to $\leq$10 autonomous steps before human intervention, with reliability cited as the primary deployment barrier. Critically, non-determinism persists even at temperature=0 \citep{ouyang2023nondeterminism, atil2024nondeterminism}, undermining reproducibility in interpreted AI approaches. Benchmarks compound this picture: HumanEval is saturated and contaminated \citep{yang2023rethinking, xu2024contamination}, contamination-resistant SWE-Bench Pro~\citep{shi2025swebenchillusion} drops to 23\%, the synthetic-to-real gap reaches 50+ percentage points \citep{rao2025synthetic}, and developers \emph{perceived} 20\% productivity gains while objective measurements showed 19\% slower completion~\citep{metr2025productivity}. Our core architectural choice, removing the LLM from the orchestration loop and confining open-ended invocation to a one-time compilation phase, directly addresses these failure modes.

\subsection{Constrained Generation and Formal Verification}
\label{sec:related-constrained}

Structured generation techniques offer complementary paths to reliability. SynCode \citep{ugare2024syncode} achieves 96\% reduction in syntax errors through grammar-constrained decoding; type-constrained decoding \citep{mundler2025typeconstrained} reduces compilation errors by $>$50\% while improving functional correctness; Outlines~\citep{willard2023outlines} and XGrammar~\citep{dong2024xgrammar} together deliver up to 100$\times$ speedup over prior constrained decoding approaches. Formal verification shows promise for critical systems: Astrogator \citep{councilman2025astrogator} verifies correct code in 83\% of cases. However, \citet{xu2024hallucination} demonstrate that structural hallucinations cannot be eliminated by larger training sets alone, positioning constrained decoding as complementary rather than complete. Our approach combines constrained generation with multi-stage validation. For healthcare specifically, \citet{neupane2025hipaa} present a HIPAA-compliant agentic AI framework integrating attribute-based access control with PHI sanitization, an approach our template-based architecture can encode as compliance-by-construction.

\section{System Architecture}
\label{sec:architecture}

The compiled AI architecture is motivated by a fundamental asymmetry in enterprise workflow automation: generating correct business logic benefits from LLM reasoning, but executing the orchestration of that logic thousands of times per day does not. Runtime agent systems conflate these two phases, invoking an LLM at every transaction. Our architecture separates them, confining LLM invocation in the control plane to a one-time compilation step. Subsequent execution runs as deterministic code, with runtime LLM invocation permitted only as bounded tool calls for narrow semantic subtasks.

This separation yields several benefits. Deterministic execution eliminates output variance and enables full auditability under this architecture. Every decision traces to a specific line of code, a property essential for CMS compliance in prior authorization and for FDA audit readiness in AI-enabled medical workflows \citep{fda2025}. Static code can be subjected to established software engineering tooling: static analysis, type checking, security scanning, and regression testing. And because the LLM exits the orchestration loop (with runtime invocation confined to bounded tool calls), the attack surface for prompt injection shrinks substantially, from one open-ended LLM call per transaction to a small number of schema-constrained calls per workflow.

\paragraph{Design Principles.} Four principles shape the architecture. \emph{Constrained Generation} limits LLM output to narrow, well-defined functions (20--50 lines); templates provide infrastructure, bounding the output space and reducing hallucination risk. \emph{Compilation over Interpretation} means generated code is validated, tested, and deployed as static artifacts. The LLM exits the open-ended orchestration loop; any runtime invocation is confined to bounded tool calls with schemas fixed at compile time. \emph{Validation as Requirement} means every artifact passes a four-stage pipeline before deployment, feasible precisely because we generate code rather than interpret configurations. \emph{Compliance by Construction} encodes regulatory constraints (HIPAA, PCI-DSS, SOC 2) directly in templates and prompt blocks, ensuring generated code inherits compliance properties by default.

\begin{figure}[ht]
\centering
\resizebox{\textwidth}{!}{%
\begin{tikzpicture}[
    node distance=0.6cm and 0.9cm,
    box/.style={rectangle, draw, rounded corners, minimum width=1.8cm, minimum height=0.7cm, align=center, font=\small},
    library/.style={rectangle, draw, dashed, minimum width=1.5cm, minimum height=0.55cm, align=center, font=\scriptsize, fill=gray!10},
    validation/.style={rectangle, draw, minimum width=1.1cm, minimum height=0.5cm, align=center, font=\scriptsize, fill=blue!10},
    arrow/.style={->, >=stealth, thick},
    label/.style={font=\scriptsize\itshape, text=gray}
]
\node[box, fill=yellow!20] (yaml) {YAML\\Spec};
\node[box, fill=orange!20, right=of yaml] (config) {Orch.};
\node[library, above=0.8cm of config] (templates) {Templates};
\node[library, right=0.1cm of templates] (modules) {Modules};
\node[library, right=0.1cm of modules] (prompts) {Prompts};
\node[box, fill=red!20, right=1.8cm of config] (llm) {LLM\\Generate};
\node[box, fill=purple!15, right=of llm] (assemble) {Assemble};
\node[validation, right=0.8cm of assemble] (sec) {Security};
\node[validation, right=0.08cm of sec] (syn) {Syntax};
\node[validation, right=0.08cm of syn] (exe) {Execute};
\node[validation, right=0.08cm of exe] (acc) {Accuracy};
\node[above=0.25cm of syn, xshift=0.3cm, font=\small\bfseries] {Validation Pipeline};
\node[box, fill=green!20, right=0.6cm of acc] (temporal) {Temporal\\Activity};
\node[below=0.6cm of assemble, font=\scriptsize, text=red!70] (regen) {Regenerate on failure};
\draw[arrow] (yaml) -- (config);
\draw[arrow] (config) -- (llm);
\draw[arrow] (llm) -- (assemble);
\draw[arrow] (assemble) -- (sec);
\draw[arrow] (sec) -- (syn);
\draw[arrow] (syn) -- (exe);
\draw[arrow] (exe) -- (acc);
\draw[arrow] (acc) -- (temporal);
\draw[arrow, gray] (templates.south) -- ++(0,-0.3) -| ([xshift=-0.2cm]llm.north);
\draw[arrow, gray] (modules.south) -- ++(0,-0.2) -| (llm.north);
\draw[arrow, gray] (prompts.south) -- ++(0,-0.3) -| ([xshift=0.2cm]llm.north);
\draw[arrow, red!70, dashed] ([yshift=-0.1cm]sec.south) |- (regen.east);
\draw[arrow, red!70, dashed] (regen.west) -| ([xshift=0.3cm]llm.south);
\node[label, below=0.05cm of yaml] {Intent};
\node[label, below=0.05cm of temporal] {Deterministic};
\node[label, above=0.05cm of llm, xshift=0.3cm] {One-time};
\end{tikzpicture}
}%
\caption{The code foundry architecture. Business intent (YAML) enters; validated Temporal activities emerge. Dashed red arrows show the regeneration loop on validation failure. The LLM runs once at generation time. Generated artifacts may include bounded LLM tool calls for steps requiring semantic understanding at runtime.}
\label{fig:architecture}
\end{figure}
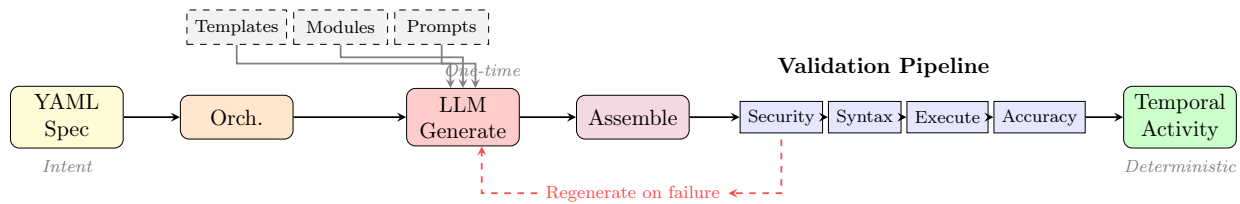

\paragraph{Component Overview.} The system takes a YAML workflow specification as input and produces a validated Temporal activity as output. An \emph{Orchestrator} receives the specification and selects appropriate templates, modules, and compliance constraints. A \emph{Template Library} provides tested code patterns for common workflow types: synchronous handlers, streaming processors, batch processors with checkpointing, and input validators with fallback logic. A \emph{Module Library} provides reusable functional capabilities including database access with connection pooling, HTTP clients with retry logic, and notification delivery. \emph{Prompt Blocks} encode domain constraints such as HIPAA data handling rules. The orchestrator assembles these components into a prompt, invokes the LLM once to generate business logic (which itself may include calls to bounded LLM tools at runtime), and passes the assembled artifact through a four-stage validation pipeline before deployment. On validation failure, the system regenerates with error context rather than deploying a broken artifact. Figure~\ref{fig:architecture} illustrates this flow; Appendix~\ref{app:algorithm} formalizes it as pseudocode.

\paragraph{Validation Pipeline.} The four stages serve distinct purposes. \emph{Security} applies static analysis (Bandit, Semgrep, custom rules) checking for SQL injection, command injection, path traversal, and secrets exposure. \emph{Syntax} performs AST parsing, type checking (mypy), and linting (ruff). \emph{Execution} runs sandboxed execution against test fixtures verifying successful completion, error handling, and output structure. \emph{Accuracy} compares outputs against golden datasets using task-specific thresholds.

\paragraph{Bounded Tool Calls and Security Architecture.} Some workflow steps require runtime judgment that cannot be fully precompiled. Our architecture supports \textit{bounded agentic invocation}: generated code may call LLMs for specific, narrow subtasks (e.g., extracting structured fields from an ambiguous clinical note) while maintaining deterministic overall flow, operating under defined schemas, fallback logic, drift monitoring, and human escalation thresholds. Beyond correctness validation, a three-gate security pipeline protects against OWASP LLM Top 10 vulnerabilities, including indirect prompt injection~\citep{greshake2023notwhat} and broader LLM-generated code security risks~\citep{veracode2025genai}: an \emph{Input Gate} (DeBERTa-v3 prompt injection detection + Presidio PII scanning), a \emph{Code Gate} (static analysis for CWE-94, CWE-89, CWE-502, CWE-78, CWE-328), and an \emph{Output Gate} (cryptographic canary token injection for prompt leakage detection). A Dual LLM pattern separates privileged and quarantined model instances. Compiled AI reduces runtime LLM invocations from one-per-transaction to zero, effectively eliminating prompt-based attack surface in the execution path.

\section{Evaluation Framework}
\label{sec:framework}

Existing benchmarks primarily measure task-level capability: code correctness or tool-use success rates. While necessary, these metrics are insufficient for evaluating production workflow automation in domains like healthcare, where economic efficiency, determinism, and reliability are more critical than expressiveness. The gap between benchmark accuracy and production viability is empirically well-documented: systems achieving 95\% benchmark accuracy may succeed in only 35\% of multi-turn production interactions \citep{salesforce2025agents}, and perceived productivity gains can diverge sharply from objective measurements \citep{metr2025productivity}.

We therefore evaluate compiled AI as a systems artifact rather than a reasoning agent, using metrics across seven categories: token efficiency, latency, consistency and determinism, reliability, code quality, validation pipeline effectiveness, and security. The token efficiency analysis centers on the break-even transaction count $n^* = \metric{GenTokens}_{\text{compiled}} / (\metric{RuntimeTokens}_{\text{per-tx, runtime}} - \metric{RuntimeTokens}_{\text{per-tx, compiled}})$. For determinism, we measure output entropy $H = -\sum p_i \log p_i$ over $N$=1,000 identical inputs in the control plane (orchestration), where compiled AI targets $H = 0$. Bounded tool calls in the data plane inherit the underlying model's variance and are evaluated separately via output validation against fixed schemas. The control-plane target contrasts with the 18--75\% output variance documented in runtime inference even at temperature=0 \citep{ouyang2023nondeterminism}. Reliability targets far exceed the 35\% agent baseline from \citet{salesforce2025agents}. Security metrics evaluate the three-gate architecture against OWASP LLM Top 10 threats. Validation pipeline metrics (\metric{FirstPassRate} per stage and \metric{OverallFirstPass}) are novel; to our knowledge no prior work reports these for LLM-generated production code. Full metric tables for all seven categories are provided as supplementary material.

\section{Experiments}
\label{sec:experiments}

We evaluate on two distinct task types to demonstrate that compiled AI generalizes beyond a single benchmark and that the architecture is not overfit to a particular input modality.

\paragraph{Task 1: Function Calling (BFCL).} The Berkeley Function-Calling Leaderboard \citep{patil2024gorilla} measures function-calling capability on 400 instances where the model must identify the correct function and extract parameters from natural language queries. This task validates the core compilation-vs-runtime trade-off in a structured setting with clean inputs. This task lies in the regime where pure deterministic compilation is feasible, since function signatures are well-typed and outputs are unambiguous.

\paragraph{Task 2: Document Intelligence (DocILE).} To stress-test the architecture on noisy, semantic inputs, we evaluate on DocILE~\citep{simsa2023docile}, 5,680 invoice documents with degraded OCR quality, heterogeneous formats across hundreds of vendors, and semantic ambiguity. DocILE uses two evaluation metrics: \emph{KILE} (Key Information Line-level Extraction), measuring exact-match accuracy on key fields such as invoice number, date, and total amount; and \emph{LIR} (Line Item Recognition), measuring accuracy on structured line items within documents. This task requires runtime semantic understanding for individual extraction steps, motivating the Code Factory variant where compiled orchestration invokes the LLM as a bounded tool.

\paragraph{Baselines.} We compare four approaches across both tasks: \emph{Direct LLM} (Claude Opus 4.5~\citep{anthropic2025opus}, one call per transaction), \texttt{LangChain} and \texttt{AutoGen} (same LLM with orchestration overhead), and \emph{Compiled AI / Code Factory} (LLM invoked once at compilation for orchestration; Code Factory additionally invokes the LLM as bounded tool calls during execution for semantic subtasks). For DocILE, we additionally evaluate a \emph{Deterministic} variant (pure regex, no LLM) to bound the non-learning approach. We chose BFCL and DocILE over agent-style benchmarks such as $\tau$-bench~\citep{yao2024taubench} because they better isolate the structured-versus-semantic input dimension central to our study. All experiments use Claude Opus 4.5 (\texttt{claude-opus-4-5-20251101}) at temperature 0, PydanticAI for structured orchestration, and a custom DSL executor for workflow definitions.

\section{Results}
\label{sec:results}

\subsection{Task 1: Function Calling (BFCL)}

\paragraph{Token efficiency and cost.} Compiled AI incurs a one-time 9,600-token generation cost and zero runtime tokens in the control plane thereafter (Table~\ref{tab:token-results}). Break-even against direct LLM occurs at $n^* \approx 17$ transactions. At 1,000 transactions, compiled AI uses 57$\times$ fewer tokens than direct LLM and 84$\times$ fewer than AutoGen. At 1M transactions/month, total cost of ownership is \$555 vs.\ \$22,000 for direct LLM, a 40$\times$ cost ratio (Table~\ref{tab:cost-results}).

\begin{table}[ht]
\centering
\caption{Token consumption comparison (BFCL, n=400). Runtime tokens reported for the function-calling task; the Code Factory variant for document intelligence is reported in Table~\ref{tab:docile-main}.}
\label{tab:token-results}
\begin{tabular}{lrrrrr}
\toprule
Method & GenTokens & Runtime/tx & Total@1K & BreakEven & CompRatio \\
\midrule
Direct LLM & 0 & 552 & 552,000 & n/a & n/a \\
LangChain  & 0 & 740 & 740,000 & n/a & 0.75$\times$ \\
AutoGen    & 0 & 805 & 805,000 & n/a & 0.69$\times$ \\
Compiled AI & 9,600 & 0 & 9,600 & $\sim$17 tx & $\sim$57$\times$ \\
\bottomrule
\end{tabular}
\end{table}

\begin{table}[ht]
\centering
\caption{Total cost of ownership at 1M transactions/month (Claude Opus 4.5: \$15/1M input, \$75/1M output tokens).}
\label{tab:cost-results}
\begin{tabular}{lrrrr}
\toprule
Method & Inference & Infra & Total TCO & CostRatio \\
\midrule
Direct LLM & \$21,500 & \$500 & \$22,000 & 40$\times$ \\
LangChain  & \$28,900 & \$500 & \$29,400 & 53$\times$ \\
AutoGen    & \$31,400 & \$500 & \$31,900 & 57$\times$ \\
Compiled AI & \$55 & \$500 & \$555 & 1$\times$ \\
\bottomrule
\end{tabular}
\end{table}

\paragraph{Latency and consistency.} Compiled AI executes at 4.5ms P50 on BFCL function-calling (vs.\ 2,004ms for direct LLM, a 450$\times$ improvement) with near-zero jitter (10.5ms vs.\ 1,123ms). It empirically achieves 100\% reproducibility and zero output entropy; runtime inference shows 95\% reproducibility due to model variance at temperature=0, consistent with \citet{atil2024nondeterminism}.

\paragraph{Reliability and validation.} Task completion is 96\% (384/400); all 16 failures are compilation-time failures, so successfully compiled BFCL workflows execute with 100\% reliability. Validation pipeline first-pass rates are 100\% through syntax and execution stages; the 96\% accuracy gate catches semantically-incorrect outputs that would otherwise deploy with 100\% execution success but wrong values, silent failures invisible to operators. Multi-stage validation converts these into detectable compile-time failures, achieving 100\% deployed reliability at the cost of a 4\% compilation failure rate (simulated ablation).

\paragraph{Security.} The prompt injection validator achieves 95.8\% recall with 100\% precision (zero false positives) across 30 adversarial inputs. The code safety gate achieves 75\% recall on 20 vulnerable code fixtures with 100\% precision (zero false positives on 20 benign workflows). Output gate canary detection is 12.5\% recall in simulation-based testing, reflecting methodology rather than production performance. Overall precision across 135 security test cases is 96.1\%.

\paragraph{Code quality.} Compiled artifacts achieve 99\% type coverage and 96\% test pass rate. Cyclomatic complexity~\citep{mccabe1976} averages 23.8 vs.\ 8 for human code, a known characteristic of LLM-generated switch-case structures \citep{gitclear2025} and an area for template refinement.

\subsection{Task 2: Document Intelligence (DocILE)}

DocILE represents the regime where individual extraction steps require semantic understanding, motivating the Code Factory variant where compiled orchestration retains determinism while invoking the LLM as a bounded tool for field-level extraction. Table~\ref{tab:docile-main} presents results on DocILE. Pure deterministic extraction (regex) runs 4,915$\times$ faster than direct LLM but achieves only 20.3\% KILE, an 80 percentage point gap driven by failures on semantic fields: customer names (10.1\% KILE) and currency amounts (9.2\% KILE), where contextual understanding is required.

Code Factory compiles orchestration deterministically and invokes the LLM as a bounded tool call for each extraction step, with prompt template, schema, and trigger conditions all fixed at compile time. It achieves competitive KILE accuracy (80.0\%, matching direct LLM) and the \emph{highest} LIR accuracy (80.4\%), while running at 2.3$\times$ lower latency than direct LLM. Combined with Pydantic output validation and automatic retry on schema violations, Code Factory demonstrates that compiled code wrapping focused LLM calls achieves parity with hand-tuned prompt engineering while enabling systematic software engineering benefits: versioning, testing, and full auditability.

\begin{table}[ht]
\centering
\caption{DocILE results. KILE: Key Information Line-level Extraction (exact-match on key fields, e.g.\ invoice number, date, total). LIR: Line Item Recognition (accuracy on structured line items). Code Factory makes compiled LLM calls; Deterministic uses pure regex.}
\label{tab:docile-main}
\begin{tabular}{llcccc}
\toprule
\textbf{Paradigm} & \textbf{Approach} & \textbf{KILE} & \textbf{LIR} & \textbf{Latency} & \textbf{LLM Calls} \\
\midrule
\multirow{2}{*}{\textit{Compiled AI}}
  & Deterministic (Regex) & 20.3\% & 59.7\% & 0.6 ms & None \\
  & \textbf{Code Factory} & \textbf{80.0\%} & \textbf{80.4\%} & \textbf{2,695 ms}$^\dagger$ & Compiled \\
\midrule
\multirow{3}{*}{\textit{Runtime AI}}
  & Direct LLM & 80.0\% & 74.5\% & 6,339 ms & Per-request \\
  & LangChain  & 80.0\% & 75.6\% & 6,207 ms & Per-request \\
  & AutoGen    & 77.8\% & 78.9\% & 13,742 ms & Per-request \\
\bottomrule
\multicolumn{6}{l}{\footnotesize $^\dagger$Median latency; mean 10,263 ms due to 10\% retry outliers (max 183s). First compile: 57s.} \\
\end{tabular}
\end{table}

Together, the two task types span the spectrum from fully structured to highly noisy semantic inputs. BFCL represents the structured end: inputs are well-typed function signatures with unambiguous correct outputs, making pure deterministic compilation feasible. DocILE represents the semantic end: invoice documents vary in layout, language, and field placement, requiring contextual understanding that a regex-based approach cannot provide. That compiled AI generalizes across both extremes through bounded tool call invocation in the semantic regime, achieving competitive accuracy on function-calling and matching Direct LLM on document intelligence, suggests the architecture is not overfit to a single input modality. The Code Factory variant provides a principled path for the semantic end of this spectrum: bounded LLM invocation as compiled code rather than ad-hoc runtime prompting.

\section{Discussion}
\label{sec:discussion}

Compiled AI is not a general replacement for runtime inference. Our results suggest a practical decision rule. Use full deterministic compilation for structured, predictable workflows where every step has a well-typed answer. Use Code Factory (compiled orchestration with bounded LLM tool calls) for mixed-content workflows where individual steps require semantic understanding but overall flow is well-specified. Use runtime LLM agents for genuinely open-ended tasks where the workflow itself must be discovered at runtime.

Across the two tasks a clear pattern emerges. BFCL represents the regime where full deterministic compilation is achievable, since function-calling is well-typed and unambiguous. DocILE represents the realistic enterprise regime, where orchestration logic is well-specified but individual steps require semantic understanding. The Code Factory variant is not a fallback but the architectural sweet spot for production workflows, which are almost always mixed. Pure deterministic compilation is the special case where every step happens to be structured. Runtime agents are the special case where every step happens to be semantic.

Healthcare administrative workflows exemplify the regime where compiled AI excels. Prior authorization, billing reconciliation, and clinical document processing are high-volume, well-specified by regulation, compliance-sensitive, latency-sensitive, and demand full auditability. \citet{moor2023foundation} survey the landscape of foundation models for medicine, and \citet{qiu2024healthcare} identify administrative workflow automation as a candidate application for LLM-based systems, and regulatory frameworks increasingly emphasize auditability \citep{fda2025}. Our prior authorization example (Appendix~\ref{app:example}) shows how compiled AI encodes insurance coverage rules as deterministic code, confining LLM invocation to the narrow step of extracting structured clinical factors from unstructured notes.

Constraining generation improves reliability by bounding the error space: the model cannot hallucinate incorrect API calls or database schemas since these come from pre-tested templates. For bounded runtime tool calls, schema validation catches structural errors but cannot detect semantic hallucination, requiring downstream monitoring. Deterministic execution also yields full observability. When accuracy degrades, we can pinpoint which code segment underperformed and trigger targeted regeneration. This positions compiled AI not as a one-shot process but as an \emph{evolutionary system}: deterministic execution with adaptive improvement, where production metrics continuously inform artifact refinement.

There is also an economic timing argument. LLM inference costs have fallen roughly 10$\times$ per year, but enterprise deployment volumes are growing faster still, driving total inference spend upward even as per-token costs decline \citep{openai2026pricing}. Architectures that amortize inference across transactions, rather than invoking a model per request, become economically compelling at scale. Compiled AI is one such architecture: generation cost is fixed at compile time, and execution cost is zero regardless of transaction volume. As LLM costs come to dominate enterprise operating budgets, compile-once-run-many approaches are not merely convenient. They are likely to become the default deployment pattern for well-specified, high-volume workflows.

\section{Conclusion and Future Work}
\label{sec:conclusion}

We studied compiled AI as a production systems design point for LLM-based workflow automation, with emphasis on the healthcare domain where reliability and auditability are regulatory requirements. Building on the compilation paradigm established in DSPy, LLM+P, and related work, we characterized compiled AI's operational properties across structured (BFCL) and semantic (DocILE, via Code Factory bounded tool calls) task regimes, and demonstrated consistent advantages in token economics (57$\times$ reduction at 1,000 transactions on function-calling), latency (450$\times$ improvement on function-calling), determinism in the control plane (100\% reproducibility for compiled orchestration vs.\ 95\% for runtime inference), and cost at scale (40--57$\times$ TCO reduction at 1M transactions/month).

The approach trades flexibility for determinism, cost efficiency, auditability, and reduced attack surface. These properties matter most in well-specified, high-volume, compliance-sensitive workflow regimes. We release our evaluation framework and benchmark suite to enable systematic comparison across the community.

\textbf{Limitations.} Several limitations warrant acknowledgment. \emph{Specification quality}: compiled AI assumes users can accurately specify workflows in YAML, and the ``specification problem'' remains fundamental, with iterative refinement often necessary. \emph{Bounded applicability}: not all workflows reduce to deterministic code, and tasks requiring genuine creativity or adaptation to novel situations may require runtime inference. Our evaluation covers two task types, and broader generalization requires further study. \emph{Generation failures}: 4\% of compilations (16/400 on BFCL) failed to produce accurate outputs, though all were syntactically valid and detectable before deployment. \emph{Bounded tool call drift}: Code Factory invocations rely on runtime LLM calls within fixed schemas. While schema validation catches structural errors, semantic drift in extracted values requires production monitoring, and our evaluation does not characterize this drift over long deployment horizons. \emph{Model dependence}: generated code quality depends on the underlying LLM, and model updates may require re-validation. \emph{Benchmark scope}: our evaluation framework has not yet been adopted by other systems, limiting cross-system comparability.

\textbf{Future Work.} Key open directions include natural language specification (moving beyond YAML), automatic workflow decomposition from high-level intent, continuous optimization through runtime accuracy metrics enabling targeted regeneration cycles, deeper characterization of bounded tool call behavior in long-running production deployments, formal verification of generated code properties, and cross-system benchmark adoption.

\bibliographystyle{plainnat}

\appendix

\section{Example: Prior Authorization Workflow}
\label{app:example}

This example demonstrates bounded agentic invocation for healthcare prior authorization, illustrating compiled AI in the healthcare regime discussed in Section~\ref{sec:discussion}: the LLM is confined to structured data extraction, while coverage decisions execute as deterministic, fully auditable code.

\begin{lstlisting}[language=Python, caption={Workflow specification for GLP-1 prior authorization}, basicstyle=\small\ttfamily]
# workflow_spec: prior_auth_glp1_v1.yaml
metadata:
  name: "GLP-1 Agonist Medical Necessity Review"
  compliance: ["HIPAA", "CMS_Decision_Timeframes"]
inputs:
  - name: "patient_chart_summary"
    type: "text"
logic_requirements:
  - step: "extract_clinical_factors"
    type: "bounded_invocation"
    schema:
      has_t2d_diagnosis: boolean
      current_a1c: float
      bmi: float
      has_step_therapy_failure: boolean
    validation:
      - "bmi > 10 AND bmi < 100"
  - step: "evaluate_coverage"
    type: "deterministic_rule"
    logic: |
      IF has_t2d_diagnosis THEN APPROVE
      ELSE IF (bmi >= 30) AND has_step_therapy_failure THEN APPROVE
      ELSE DENY
\end{lstlisting}

\begin{lstlisting}[language=Python, caption={Compiled artifact isolating probabilistic extraction from deterministic decisions}, basicstyle=\small\ttfamily]
@workflow.defn
class PriorAuthWorkflow:
    @workflow.run
    async def run(self, input_data: WorkflowInput) -> AuthResult:
        # PHASE 1: BOUNDED TOOL CALL
        clinical_data = await workflow.execute_activity(
            "extract_clinical_data_activity",
            args=[input_data.chart_text],
            retry_policy=RetryPolicy(maximum_attempts=3)
        )
        # PHASE 2: HALLUCINATION VALIDATION
        if clinical_data.a1c and not (3.0 < clinical_data.a1c < 20.0):
            return await self.trigger_human_review("Suspicious A1C")
        # PHASE 3: DETERMINISTIC DECISION
        decision, reason = "DENIED", "Does not meet criteria"
        if clinical_data.has_t2d_diagnosis:
            decision, reason = "APPROVED", "Type 2 Diabetes Diagnosis"
        elif (clinical_data.bmi >= 30.0 and
              clinical_data.has_step_therapy_failure):
            decision, reason = "APPROVED", "BMI>=30 + Step Therapy"
        # PHASE 4: AUDIT TRAIL
        await workflow.execute_activity("log_decision_event",
            {"decision": decision, "logic_version": "v1.0"})
        return AuthResult(status=decision, reason_code=reason)
\end{lstlisting}

This example demonstrates three properties: (1) \emph{Safety Sandwich}: the probabilistic model is constrained between input validation and deterministic logic \citep{dalrymple2024guaranteed}; (2) \emph{Auditability}: decisions trace to specific code lines, enabling full auditability under this architecture; (3) \emph{Token Economics}: coverage rules are compiled once rather than sent per-request.

\section{Benchmark Suite}
\label{app:benchmark}

We release our benchmark suite at \url{https://github.com/XY-Corp/Compiled-AI}, including 540 YAML workflow specifications, golden outputs for accuracy validation, measurement scripts for all seven metric categories, and baseline implementations.

\section{Code Foundry Generation Algorithm}
\label{app:algorithm}

Algorithm~\ref{alg:generation} formalizes the generation-and-validation procedure described in Section~\ref{sec:architecture}. The procedure takes a workflow specification, selects templates and modules, invokes the LLM once to generate code, assembles the artifact, and runs the four-stage validation pipeline (Security, Syntax, Execution, Accuracy) with regeneration on failure.

\begin{algorithm}[H]
\caption{Code Foundry Generation Process}
\label{alg:generation}
\begin{algorithmic}[1]
\Require Workflow specification $S$, Template library $T$, Module library $M$
\Ensure Validated Temporal activity $A$
\State $t \gets \textsc{SelectTemplate}(S, T)$
\State $m \gets \textsc{SelectModules}(S, M)$
\State $p \gets \textsc{AssemblePrompt}(S, t, m)$
\State $code \gets \textsc{LLMGenerate}(p)$ \Comment{One-time generation}
\State $A \gets \textsc{Assemble}(t, m, code)$
\For{$stage \in \{\text{Security}, \text{Syntax}, \text{Execution}, \text{Accuracy}\}$}
    \State $result \gets \textsc{Validate}(A, stage)$
    \If{$result = \text{FAIL}$}
        \State $code \gets \textsc{Regenerate}(p, result.errors)$
        \State $A \gets \textsc{Assemble}(t, m, code)$
    \EndIf
\EndFor
\State \Return $A$ \Comment{Deterministic artifact}
\end{algorithmic}
\end{algorithm}

\end{document}